\documentclass[twocolumn,11pt]{article}
\usepackage[left=1.7cm,right=1.7cm,bottom=2.4cm,top=2.1cm]{geometry}
\usepackage{amsmath,graphicx,amsbsy,bbm}
\usepackage{multicol,amsfonts,array}
\usepackage{flushend}
\title{Memory Requirement Reduction of Deep Neural Networks Using Low-bit Quantization of Parameters}
\author{Niccol\`{o} Nicodemo$^{\dagger}$, Gaurav Naithani$^{\ddagger}$, Konstantinos Drossos$^{\ddagger}$,\\Tuomas Virtanen$^{\ddagger}$, and Roberto Saletti$^{\dagger}$ \vspace{9pt}\\
$^{\dagger}$\{n.nicodemo1@studenti., r.saletti@\}unipi.it,\\University of Pisa, Italy\\
$^{\ddagger}$\{firstname.lastname\}@tuni.fi,\\Audio Research Group, Tampere University, Finland
}
\date{}
\begin{document}
\twocolumn[
\maketitle
  \begin{@twocolumnfalse}
    \maketitle
    \begin{abstract}
    \normalsize
    Effective employment of deep neural networks (DNNs) in mobile devices and embedded systems is hampered by requirements for memory and computational power. This paper presents a non-uniform quantization approach which allows for dynamic quantization of DNN parameters for different layers and within the same layer. A virtual bit shift (VBS) scheme is also proposed to improve the accuracy of the proposed scheme. Our method reduces the memory requirements, preserving the performance of the network. The performance of our method is validated in a speech enhancement application, where a fully connected DNN is used to predict the clean speech spectrum from the input noisy speech spectrum. A DNN is optimized and its memory footprint and performance are evaluated using the short-time objective intelligibility, STOI, metric. The application of the low-bit quantization allows a 50\% reduction of the DNN memory footprint while the STOI performance drops only by 2.7\%. 
    
    \vspace{6pt}
\end{abstract}
\vspace{18pt}
\end{@twocolumnfalse}
]
\section{Introduction}
\label{sec:intro}
Field programmable gate arrays (FPGAs) are widely used in mobile devices as they allow for the design of highly efficient systems, with low-latency and low-power requirements. FPGAs are particularly useful for speeding up signal processing by using specific designed hardware (called hardware accelerators) to be run in parallel with main CPUs, usually embedded in the FPGA itself.
Deep neural networks (DNNs) often set the state-of-the-art in many signal processing tasks, e.g., speech separation~\cite{luo2019conv, wang2018supervised, wang2018alternative}, speech recognition~\cite{chiu2018state}, etc. However, memory footprint, memory bandwidth requirements, and the associated power consumption of DNNs are a issue to be solved for the deployment of a DNN on an FPGA. 

Two main approaches have been used to decrease the memory requirements for neural networks: i) \textit{changing the architecture} of the network in order to reduce the parameter number, and ii) \textit{quantizing the parameters} of the network to directly reduce the amount of memory needed for storing them (i.e. reducing the memory footprint) and the memory bandwidth needed to read them. The first approach involves methods like \textit{parameter pruning and sharing} \cite{han2015deep}, i.e., removing redundant weights or layers, \textit{knowledge distillation} \cite{bucilua2006model}, i.e., retrieving a smaller network from a pretrained bigger one, and the use of \textit{low-rank factorization} \cite{sainath2013low} or specific convolutional filters \cite{cheng2017survey}. All these methods produce networks with less computational needs but require a modification in the architecture of the network itself. This is less desirable from the perspective of hardware deployment, since each change in the architecture affects the hardware design and may mean to design a specific hardware for a specific architecture. Additionally, all the above mentioned methods require the optimization of the new DNN architecture. 

The second approach may lead to quantizing the network parameters from floating-point (e.g. 32-bit) to a $n$-bit fixed-point representation. In FPGAs the parameters of a DNN are usually stored in external memories (i.e. flash memories). The access time for a flash memory can be a bottleneck and severely slow down the corresponding calculations for the DNN. Reducing the number of bits needed to store each DNN parameter reduces memory requirements and improves the execution speed. Furthermore, smaller, slower, and cheaper memories can be used by employing low-bit fixed point arithmetic, resulting in a reduction of the power consumption also~\cite{dally2015high}. However, the parameter quantization can lead to a degradation of the DNN performance and very poor results if too few bits are used (i.e. less than 8)~\cite{choukroun2019low}. Several quantization strategies have been tried like normalization~\cite{solovyev2018fpga}, uniform and non-uniform quantization for different ranges of values~\cite{park2018value}, using Minimum Mean Squared Error~\cite{choukroun2019low}, weights clipping and bias correction~\cite{banner2018post}, and per-channel or per-layer different scaling~\cite{banner2018post,qiu2016going,judd2015reduced}. Mixed approaches came up too, like \textit{binarized neural network}~\cite{courbariaux2016binarized}, in which weights and activations are forced to -1 and +1 values, requiring a specific architecture and a specific training for the network. 

In this paper we consider the quantization of the weights of a DNN, we focus on the use-case of FPGAs, and we propose a low-bit quantization method based on the non-uniform and dynamic quantization methods~\cite{park2018value,qiu2016going,articleWeightsQuantization,liss2018efficient,HwSwCodesign}. Our approach distinguishes itself from earlier similar works by introduction of a virtual bit shift (VBS) scheme that allows for dynamically adjusting parameter representation for parameter ranges within the same layer as well as for different layers. VBS mitigates the drawbacks of fixed-point quantization scheme and increases the accuracy thereby reducing performance loss. Our method encodes the parameters of the DNN employing a probabilistic-based and hardware-oriented approach, using codes that can be stored in slow, external memories, while the actual values can be kept in FPGA-mapped lookup tables (LUT). Specifically, we apply a quantization which stores 4-bit codes of the parameters in external memory, thus reducing the memory footprint up to 50\%, if compared to an 8-bit fixed point representation of the parameters. The quantization technique is applied to a speech separation task, achieving the aforementioned footprint reduction with a performance reduction of only 2.7\% in terms of STOI. Furthermore, using 4-bit codes reduces the bandwidth requirement too. In fact, halving the bitwidth of the stored weights halves the bandwidth of the memory accesses, which often represents a bottleneck of the whole system. 
\section{Proposed Quantization Method}\label{sec:proposed-method}
Our method consists in taking as an input the set of parameters $\Theta$ of a deep neural network (DNN), quantizing them with fixed point values of $m$-bit width 
by applying a non-uniform quantization, and then encoding the $m$-bit values using codes of a $n$-bit lookup table (LUT) that associates the $n$-bit wide codes to the $m$-bit wide values. The $n$-bit wide codes are stored in an external, slow, memory and the fixed-point $m$-bit wide values of the parameters of the network are kept in the FPGA memory and are retrieved using the LUT. 
\subsection{Quantization of parameters $\boldsymbol{\Theta}$}
\begin{figure}[t]
	\centering
	\includegraphics[width=0.8\columnwidth]{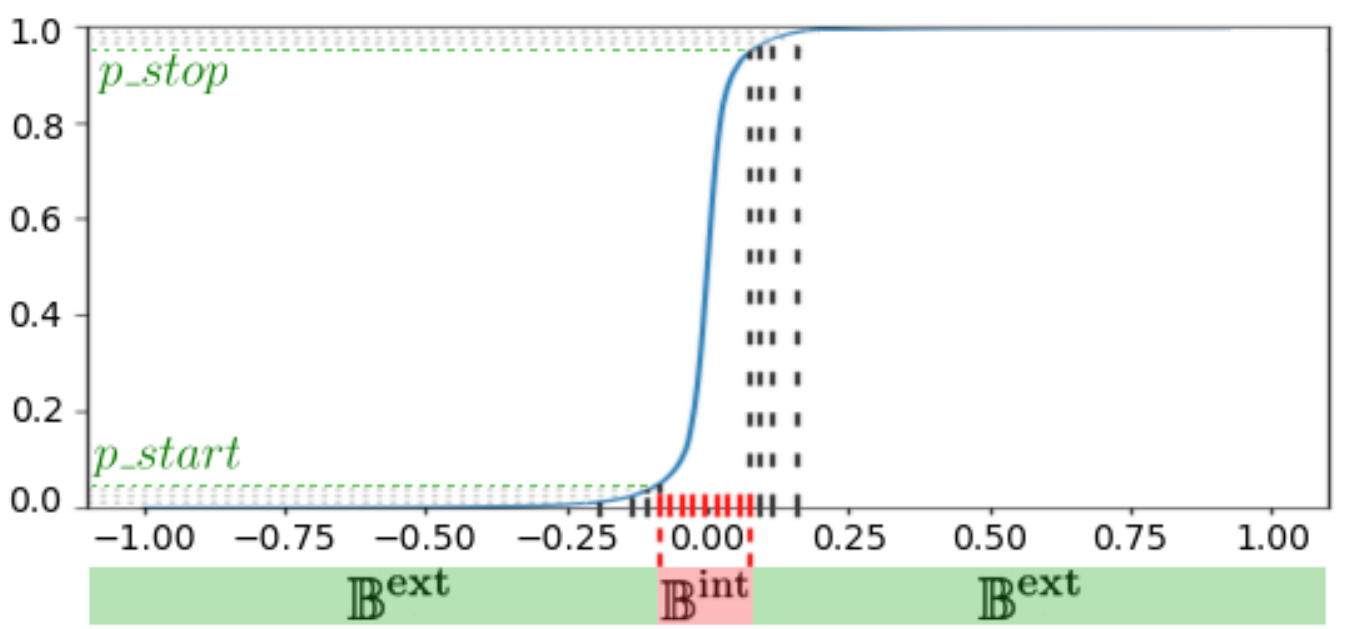}
	\caption{Cumulative distribution and partitions}
	\label{fig:quant_plot}
	\end{figure}

Any quantization scheme that converts the DNN parameters $\Theta$ from floating to fixed point values leads to quantization errors and subsequent performance losses. The aim of any such scheme is to reduce this error to a minimum. As $\Theta$ is generally non-uniformly distributed, it seems appropriate to use a non-uniform scheme. Given $\Theta$, its range can be expressed as $A=[a^{l}, a^{h}]$, where $a^{l}\leq\theta\leq a^{h},\,\theta\in\Theta$. We can use an $n$-bit encoding scheme for quantizing the range $A$ into discrete intervals, resulting into $2^n$ intervals $\mathbb{B}=\{B_{i}\}_{i=1}^{2^n}$, with $B_{i}=[b^{l}_{i},b^{u}_{i}]$, and $b^{u}_{i-1}<b^{l}_{i}<b^{u}_{i}$. For the purpose of illustration, we will from now on consider the cumulative distribution of parameters $\phi$, shown in Figure~\ref{fig:quant_plot}, as example of a feedforward network where the parameter values are clamped to the range $[-1, 1]$. The quantization error corresponding to each interval is directly related to the interval $\Delta_i=b^{u}_{i}-b^{l}_{i}$. We divide $A$ into two partitions: $\mathbb{B}^{\text{int}}\subset\mathbb{B}$ termed as the \emph{internal} partition, where the parameters are densely concentrated, and $\mathbb{B}^{\text{ext}}\subset\mathbb{B}$, termed as the \emph{external} partition, where the parameters are sparsely concentrated. The interval span in the $\mathbb{B}^{\text{ext}}$ is larger than the interval span in $\mathbb{B}^{\text{int}}$, i.e. $B_{i}\in\mathbb{B}^{\text{int}},\,B_{e}\in\mathbb{B}^{\text{ext}}\Rightarrow\Delta_{i}<\Delta_{e}$, and $\mathbb{B}^{\text{int}}\cap\mathbb{B}^{\text{ext}}=\emptyset$.

We use uniform quantization in $\mathbb{B}^{\text{int}}$ and non-uniform quantization in $\mathbb{B}^{\text{ext}}$. We can define the ratio of number of intervals in the internal and external partitions $R_{\text{B}}=|\mathbb{B}^{\text{int}}|/|\mathbb{B}^{\text{ext}}|$, where $|\cdot|$ is the number of elements in a set, and the probability values $p_{\text{start}}$ and $p_{\text{stop}}$ denoting the lower and upper boundaries of $\mathbb{B}^{int}$, respectively. We define the number of intervals $|\mathbb{B}^{\text{int}}|$ and $|\mathbb{B}^{\text{ext}}|$ as

\begin{align}
    |\mathbb{B}^{\text{ext}}| &= \left\lfloor{\frac{2^n}{1 + R_{\text{B}}}}\right\rfloor + c\text{, and}\\
    |\mathbb{B}^{\text{int}}| &= 2^n - |\mathbb{B}^{\text{ext}}|\text{, \hspace{2mm} where}\\
    c &= \begin{cases}
        1, & \text{if }\left\lfloor{\frac{2^n}{1 + R_{\text{B}}}}\right\rfloor\text{ is odd}\\[6pt]
        0, & \text{if }\left\lfloor{\frac{2^n}{1 + R_{\text{B}}}}\right\rfloor\text{ is even.}
    \end{cases}
\end{align}
\noindent

For the external partition, we uniformly split the range of $\phi$ and invert it back to get the set of intervals $B_i^{ext}$ as

\begin{equation}
    B^{\text{ext}}_i = \left[ \phi^{-1}(i \cdot \Delta^{\phi}_i), \phi^{-1}([i+1] \cdot \Delta^{\phi}_i)~\right) ,
\end{equation}
\noindent

where $\Delta^{\phi}_i=\frac{2 \cdot p_{\text{start}} }{|\mathbb{B}^{\text{ext}}|}$ is the interval span in the range of $\phi$ and hence the corresponding $\Delta_i$ is non uniform. Finally, we uniformly divide $\mathbb{B}^{\text{int}}$ with step $\Delta_i = \frac{\phi^{-1}(p_{\text{stop}}) - \phi^{-1}(p_{{\text{start}}})}{|\mathbb{B}^{\text{int}}|}$ , thereby obtaining a set of intervals $B_i^{int}$ as

\begin{equation}
\begin{split}
B^{\text{int}}_i = \left[ \phi^{-1}(p_{\text{start}}) + i \cdot \Delta_i,\right. \\ ~~~~~\left. \phi^{-1}(p_{{start}}) +\left( i+1 \right) \cdot \Delta_i \right) .
\end{split}
\end{equation}
\noindent

For any such interval $B_i$, the quantized level $\tilde{\theta}_{i}$ can be computed as the $m$-bit quantized mean of the parameters lying in the $B_{i}$ interval as,

\begin{align}
	&\tilde{\theta}_i = {\frac{\sum_{\theta\in\Theta}\mathbbm{1}_{\theta\in B_{i}}\theta}{\sum_{\theta\in\Theta}\mathbbm{1}_{\theta\in B_{i}}}}\Big \arrowvert_{m\text{-bit}}\text{, \hspace{2 mm} where}\label{eq:avg_val} \\
	&\mathbbm{1}_{\Xi}=\begin{cases}
	    1,&\text{if~~}\Xi,\\
	    0,&\text{otherwise,}
	\end{cases}
\end{align}
\noindent

and $\cdot\arrowvert_{m\text{-bit}}$ means the $m$-bit representation. Using $\tilde{\theta}_i$ from Eq.~\eqref{eq:avg_val}, we can quantize the parameters $\Theta$ of a DNN with an $m$-bit representation. The finite amount of values assumed by $\tilde{\theta}_i$, enables the reduction of the memory word length from $m$ to $n$, with $n<m$. This is achieved through a lookup table (LUT) which stores the relationship between the $n$-bit code and its corresponding $m$-bit value and partition (i.e. external or internal). An example of such a LUT is shown in Table~\ref{tab:tab1}.

\begin{figure}[t!]
	\centering
	\includegraphics[width=.55\columnwidth,trim={0.65cm .2cm .7cm .7cm},clip]{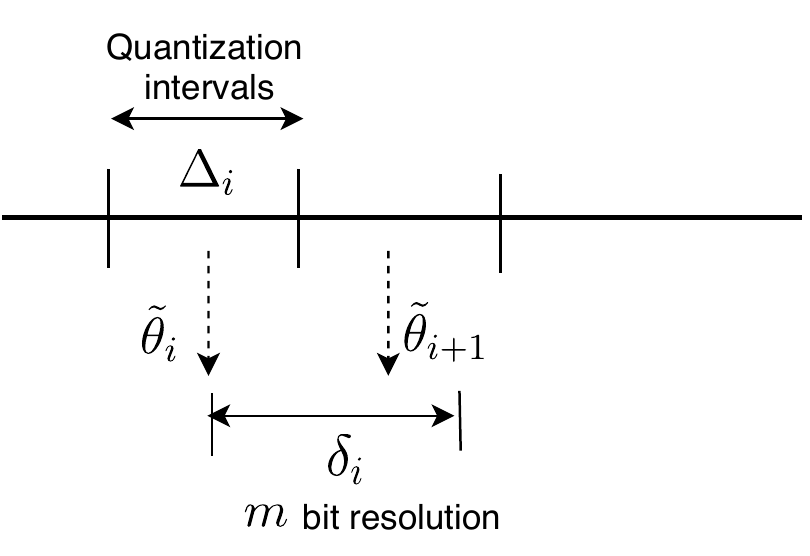}
	\caption{Example with the resolution of quantization $\Delta_i$ being smaller than the resolution of $m$-bit encoding $\delta_i$. } 
	\label{fig:resolution}
\end{figure}

\begin{table}[t!]
	\centering
	\caption{LUT of the $n$-bit code, the $m$-bit wide $\tilde{\theta}_i$, and the corresponding partition. }
	\label{tab:tab1}
	\bigskip
    \begin{tabular}{ l@{\hspace{7ex}} c@{\hspace{7ex}} c}
	Code $\vert_{10}$ & $\tilde{\theta}_i$ & Partition \\
	\hline
	0000 & -0.3359375 & ext\\
	0001 & -0.15625 & ext\\
	0100 & -0.0703125 & int\\\
	\vdots & \vdots & \vdots
\end{tabular}
\end{table}

\subsection{Virtual bit shift}
The encoding scheme using $\mathbb{B}$ is heavily dependent on the choice of the parameters $p_{\text{start}}$ and $p_{\text{stop}}$ (boundaries of $\mathbb{B}^{\text{int}}$). The smallest number that can be represented using signed $m$-bit encoding, i.e., the resolution of the encoding scheme $\delta_i$, is $2^{-(m-1)}$ and hence is dependent on the bit-width. If the span from $p_{\text{start}}$ to $p_{\text{stop}}$ is very narrow, we may end up to an interval span $\Delta_i$ smaller than $\delta_i$. In that case, the adjacent intervals will map to the same $m$-bit value as the resolution of encoding scheme $\delta_i$ is less accurate than the interval partition $\Delta_i$ . The Figure \ref{fig:resolution} depicts such a situation, where $\tilde{\theta}_i$ and $\tilde{\theta}_{i+1}$ are the values for quantized parameters corresponding to the $i^{th}$ and $i+1^{th}$ interval, respectively, and that share the same $m$-bit representation. To avoid this there should be, $\delta_i < \Delta_i$.

We propose a different quantization scheme for $\mathbb{B}^{\text{int}}$. Since $\delta_i$ depends on the bit-width of the encoding, a higher resolution can in principle be achieved by quantizing $\tilde{\theta}_i$ using $m+k$ bits. Since for all $\tilde{\theta}_i \in \mathbb{B}^{\text{int}}$, we have $\tilde{\theta}_i \leq max(abs(\phi^{-1}(p_{start}), \phi^{-1}(p_{stop})))$, and variable $k\in\mathbb{N}$ can be found so that $\tilde{\theta}_i < 2^{-k}$. This implies that the $k$ most significant bits will contain either zeros or the sign bit and can be considered redundant for storage purposes. The sign bit can be stored in the $n$ bit indexing code itself. Storing only $m$ least significant bits from a $m+k$-bit representation of $\tilde{\theta}_i$ can be thought of as shift of $k$ bits to the left which implies multiplication by $2^k$ in binary arithmetic. Let us denote $m$ least significant bits for $\tilde{\theta}_i$ as $\tilde{\theta}_i^{m}$, so that we have,

\begin {equation}
\label{eq:bit_shift}
\tilde{\theta}^m_i = \tilde{\theta}_i \cdot 2^k \text{.}
\end{equation}
\noindent

We can store $\bar{\theta}_i^m$, from which the actual parameter values $\tilde{\theta}_i$ can be retrieved using Eq. (\ref{eq:bit_shift}). Basically we perform a range adjustment by virtual bit shift of actual parameter values. An example of the same is shown in Table \ref{tab:tab2}. The resolution error can now be avoided by observing a lesser stringent condition than before, namely, $\delta_i^{m+k} < \Delta_i$, where $\delta_i^{m+k}$ is the resolution of $m+k$ bit encoding. Thus absolute values, signs and representation range information can be stored in the same code and conversion table mapped in a FPGA-embedded LUT. A $n$-bit quantization is obtained in which actual parameter values are not bounded to uniformly quantized values, but can be chosen in a proper way in order to reduce errors.

\begin{table}[t!]
    \centering
    \caption{An example of LUT with virtual bit shift for $n$ = 4, $m$ = 8, and $k$ = 4.}
    \label{tab:tab2}
    \bigskip
    \resizebox{.99\columnwidth}{!}{
        \begin{tabular}{cccc}
        Code & value & $ \text{binary value}^{12bit}$ & LUT value \\
        \hline
        0100 & 0.02099609  & .000001010110 & 01010110
        \end{tabular}}
\end{table}

\section{DNN-based speech enhancement}
We apply the proposed quantization scheme on a speech enhancement task using a feedforward DNN. Input noisy mixtures are represented using the magnitude
short-time Fourier transform (STFT) and then scaled in order to properly calculate their magnitude and phase by using a coordinate rotation digital computer (CORDIC) algorithm~\cite{volder1959cordic} based on integer arithmetic. $N$ frames, $\{\tilde{\mathbf{x}}_{t-N+1}, \tilde{\mathbf{x}}_{t-N}, \tilde{\mathbf{x}}_{t-N-1} \ldots,\tilde{\mathbf{x}}_{t}\}$ of these features are first stacked together and then fed to the DNN to estimate denoised/clean speech magnitude spectrum $\mathbf{x}_{t}$. The stacking of features is done to allow the DNN to implicitly model temporal dependencies. The CORDIC algorithm is then applied again on $\mathbf{x}_{t}$ to restore phase
information extracted from the mixture features. The values
thus obtained are scaled back and converted back to time domain speech via inverse fast Fourier transform (IFFT) and overlap-add.

\section{Evaluation}\label{sec:evaluation}
For evaluation, synthetic mixtures are created using Wall Street Journal (WSJ0) dataset for speech and TUT Acoustic scenes 2016 development dataset \cite{TUT_scene} for noise. The latter consists of sound recordings from 15 real-world environments, e.g., cafe, train, metro station, etc. A random speech signal is selected and an equal-length noise segment is sampled from the noise signal. The training and validation data consist of about 12,000 (around 20 hours) and 5000 mixtures ( around 8 hours), respectively. Similarly, the test data consists of about 2800 mixtures (around 5 hours). The speech and noise signals are mixed with a randomly chosen signal to noise ratio (SNR) from the set \{0, 5\} dB. The native sampling rate for noise signals is 44.1 kHz which is down-sampled to 8 kHz, the native sampling rate of WSJ0 audio.The short term objective intelligibility (STOI) \cite{taal2010short} metric is used as a measure of intelligibility of enhanced speech. 

The STFT features are extracted with Hann window of 128 sample (16 ms) with 50\% overlap. Eight input frames are stacked and fed to a two-layer feedforward network with 256 and 129 neurons in input and hidden layer, respectively. The rectified linear unit is used as non-linearity for each layer. The Adam optimization \cite{kingma2014adam} with default parameters is used. For training networks, PyTorch \cite{paszke2017automatic} library is used, and for audio processing, Librosa \cite{librosa} library is used. Since our focus is fixed point arithmetic devices, the network weights and biases are clamped to the range (-1, +1) in order to avoid overflow and reduce the number of bits needed for correct numeric representation in network's operations. The DNN weights have been quantized using $n=4$, $m=8$, and $k_{1^{st}layer}=3, $ $k_{2^{nd}layer}=2, $ obtained by choosing $ratio=1$, $p_{\text{start}} = 0.04$ and $p_{\text{stop}}= 0.96$. The $ratio$, $p_{\text{start}}$ and $p_{\text{stop}}$ have been chosen empirically after optimizing for the test data. The biases used are the 8-bit-uniform quantized values.

\begin{table}[t]
    \centering
    \caption{Evaluation results in terms of STOI and memory reduction. $m$-U is for $m$-bit uniform and $m$-NU for $m$-bit non-uniform quantization. Memory reduction is with respect to 8 bit fixed-point uniform quantization as reference.}
    \label{tab:res1}
    \bigskip
    \resizebox{.99\columnwidth}{!}{
        \begin{tabular}{llll}
        \multicolumn{2}{c}{\textbf{Quantization scheme}} & & \\
        \cline{1-2}
        \textbf{1\textsuperscript{st} layer} & \textbf{2\textsuperscript{nd} layer} & \textbf{STOI/STOI Loss} & \multicolumn{1}{m{2.2cm}}{\textbf{Memory usage (bytes) /reduction}}\\
        \hline
        8-U & 8-U & 0.87/-0.002 (00.2\%) & 297216/~~~~-- \\
        4-U & 4-U & 0.63/-0.246 (28.2\%) & 148608/~-50.0\% \\
        4-NU & 4-NU & 0.85/-0.024 (02.7\%) & 148608/~-50.0\% \\
        4-NU & 8-U & 0.86/-0.016 (01.8\%) & 165120/~-44.4\% \\
        \multicolumn{2}{l}{No quantization} & 0.87/~~~~-- & \\
        \multicolumn{2}{l}{Mix} & 0.84/~~~~-- & 
        \end{tabular}
    }
\end{table}

\begin{figure}[t!]
	\centering
	\includegraphics[width=0.99\columnwidth]{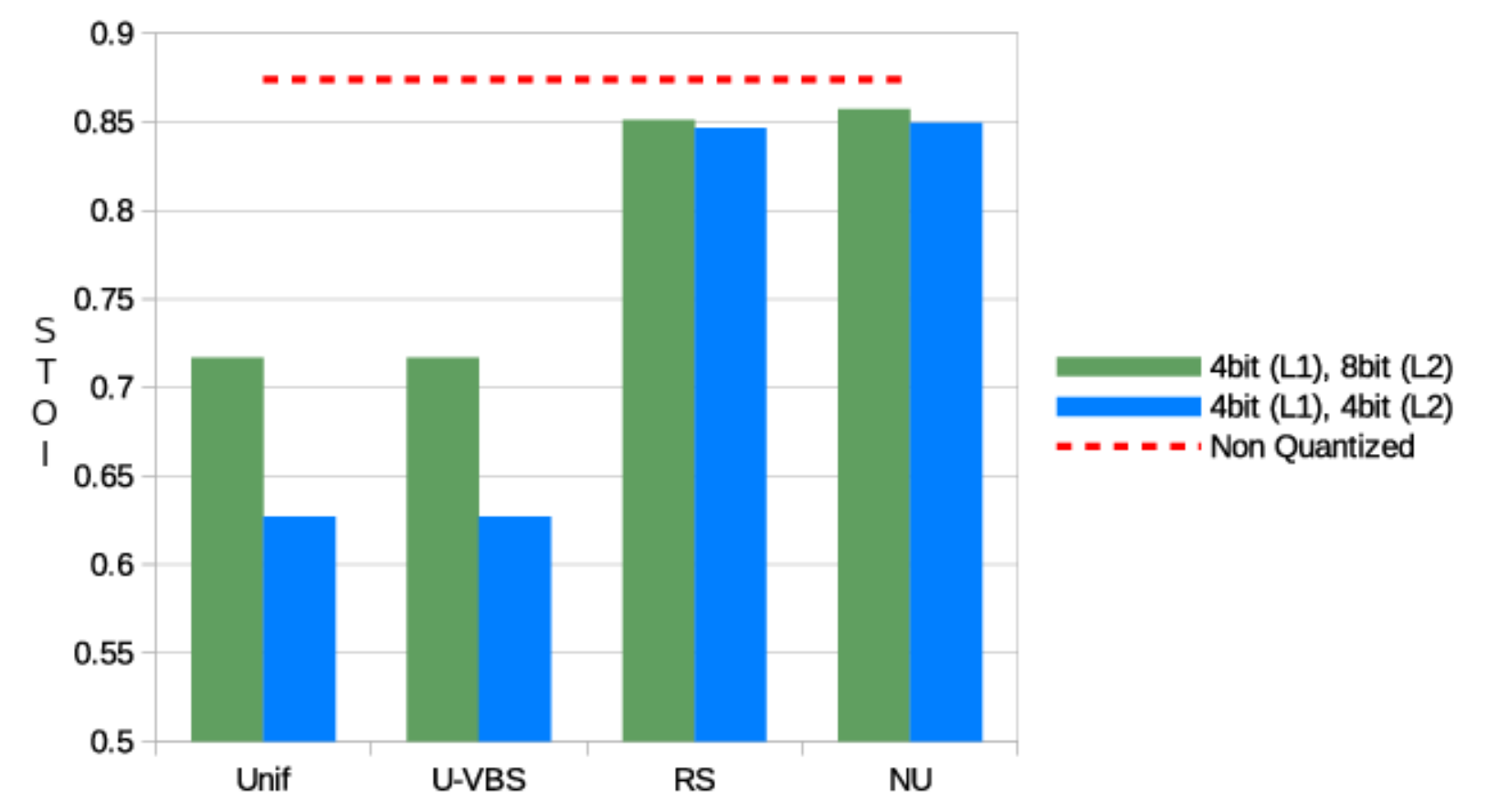}
	\caption{Comparison in terms of STOI performance of different quantization approaches being swept on first/second DNN layer keeping the other with 8-bit uniform quantization.}
    \label{img:8bit_comp}
\end{figure} 

\section{Results}
Table \ref{tab:res1} compares the STOI values obtained with different approaches over 2800 noisy samples.
8-bit-uniform quantization gives good results with a very small degradation in STOI (0.21\%) compared to non-quantized network, while 4-bit-uniform quantization led to a drastic fall in the performance, obtaining a result that is less intelligible than even the input noisy signal. On the other hand, the 4-bit non-uniform quantization proposed in this paper yields better STOI than noisy mixtures and only 2.7\% worse as compared to the non-quantized network and halving the memory footprint in comparison to the 8-bit uniform quantization while simultaneously decreasing the memory bandwidth requirement. 

\begin{figure}[t!]
	\centering
	\includegraphics[width=0.9\columnwidth]{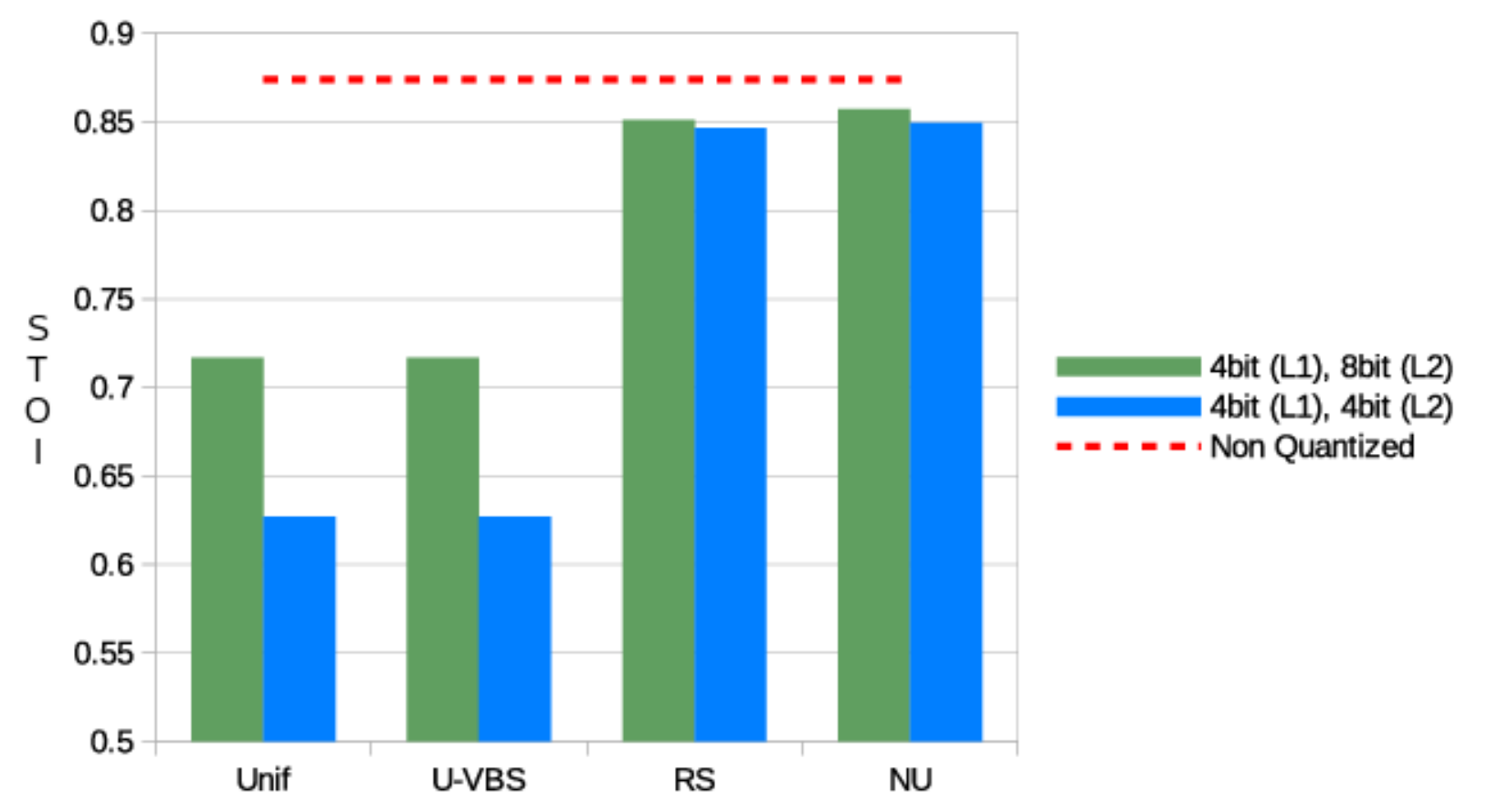}
	\caption{Comparison in terms of STOI performance of different quantization approaches with 4/8-bit quantization for first and second layer.}
  \label{img:var_comp}
\end{figure}

Figure \ref{img:8bit_comp} compares the different approaches by using 8-bit quantization (uniform and non uniform) for different layers, and how the proposed quantization scheme consisting of range split (RS) and virtual bit shift (VBS) affects the performance. For each simulation, one of the two layer is kept at 8-bit uniform quantization while the other is swept between the following four approaches: uniform quantization (U), uniform quantization with virtual bit shift (UVBS), range split (RS), and range split with virtual bit shift (RSVB). It can easily be noticed that for the first layer sweep, when no VBS is used, how the performance suffers as $\delta_i > \Delta_i$. Figure~\ref{img:var_comp} shows the effect of these approaches for two cases: 4-bit quantization for both layers, and, 4-bit for the first and 8 bit for the second layer.

\section{Conclusions}
This work proposes a low-bit quantization method inspired by the companding approach that allows the achievement of a good trade-off between performance and resource requirements in a hardware implementation of a DNN and is thus very appealing for FPGA applications. The method does not require any change or pruning of the network, so no retrain is needed. The case studied shows a two-layer feed-forward neural network, from which it emerges that a dramatic reduction of the memory requirements is obtained (50\%) with only a slight reduction of the performance. Further research should concern the application of the method to deeper networks and the usage of non symmetrical range split or of a custom multiplying architecture for the weighting of the input values.

\section*{Acknowledgement}
The authors wish to acknowledge CSC-IT Center for Science, Finland, for computational resources.

\bibliographystyle{IEEEtran}
\bibliography{strings,refs}
\end{document}